# CLUSTERING AND LARGE-SCALE STRUCTURE WITH THE SDSS

Neta A. Bahcall

Princeton University Observatory

Princeton, NJ 08544

## Abstract

The Sloan Digital Sky Survey (SDSS) will provide a complete imaging and spectroscopic survey of the high-latitude northern sky. The 2D survey will image the sky in five colors and will contain nearly 5 x $10^7$ galaxies to g ~ $23^m$. The spectroscopic survey will obtain spectra of the brightest $10^6$ galaxies, $10^5$ quasars, and $10^{3.5}$ rich clusters of galaxies (to g~18.3-19.3$^m$, respectively). I summarize some of the science opportunities that will be made possible by this survey for studying the clustering and large-scale structure of the universe.

The survey will identify a complete sample of several thousand rich clusters of galaxies, both in 2D and 3D - the largest automated sample yet available. The extensive cluster sample can be used to determine critical clustering properties such as the luminosity-function, velocity-function, and mass-function of clusters of galaxies (a critical test for cosmological models), detailed cluster dynamics and $\Omega$(dyn), the cluster correlation function and its dependence on richness, cluster evolution, superclustering and voids to the largest scales yet observed, the motions of clusters and their large-scale peculiar velocity field, as well as detailed correlations between x-ray and optical properties of clusters, the density-morphology relation, and cluster-quasar associations. The large redshift survey, reaching to a depth of $\geq$600h$^{-1}$ Mpc, will accurately map the largest scales yet observed, determine the power-spectrum and correlation function on these large scales for different type galaxies, and study the clustering of quasars to high redshifts (z$\gtrsim$4). The implications of the survey for cosmological models, the dark matter, and $\Omega$ are also discussed.

## 1. Introduction

Large sky surveys, especially in three dimensions, are vital for the study of clustering and large-scale structure of the universe. The surveys must cover huge volumes in order to investigate large-scale structure. They can provide accurate, systematic, and complete data bases of galaxies, clusters, superclusters, voids, and quasars, which can be used to study the universal structure and its cosmological implications. The surveys will allow both broad statistical studies of large samples as well as specific detailed studies of individual systems.

In this paper, I summarize some of the fundamental contributions that large scale surveys will offer for the study of clustering and large-scale structure. I will concentrate on the planned Sloan Digital Sky Survey (York *et al*. 1993); however most of the discussion is applicable in general to large-scale imaging and spectroscopic surveys.

The Sloan survey will make a powerful contribution to the study of large scale structure because of the survey's size, uniformity, and the high quality of its photometric and spectroscopic data. These characteristics will allow galaxy clustering in the present-day universe to be measured with unprecedented precision and detail. Completeness, precision and detail are keys to addressing the important physical questions, including tracing the geometrical organization of our universe and its quantitative measures, the determination of the mass density of the universe, and constraining cosmological models of galaxy formation and evolution. The breadth and quality of the Sloan Survey data make it capable of addressing a wide range of questions about clustering and about large scale structure, including questions we have not yet thought to ask.

## 2. The Sloan Digital Sky Survey

The Sloan Digital Sky Survey (SDSS) is a project which will produce a complete photometric and spectroscopic survey of half the northern sky ($\pi$ sterradians). The photometric survey will image the sky in five colors (u, g, r, i, z), using a large array of thirty $2048^2$ pixel CCD chips; the imaging survey will contain nearly $5 \times 10^7$ galaxies to g ~

$23^m$, a comparable number of stars, and about $10^6$ quasar candidates (selected on the basis of the five colors). The imaging data will then be used to select the brightest $\sim 10^6$ galaxies and $\sim 10^5$ quasars for which high resolution spectra will be obtained (to g $\sim 18.3^m$ and $19.3^m$, respectively) using 600 fibers on two high resolution (R=2000) double-spectrographs. Both the imaging and spectroscopic surveys will be done on the same 2.5-meter wide-field (3° FOV) special purpose telescope located at Apache Point, NM. The imaging survey will also be used to produce a catalog in five colors of all the detected objects and their main characteristic parameters. In the southern hemisphere the SDSS plans to image repeatedly a long and narrow strip of the sky, $\sim$ 100° x 3°, reaching $2^m$ fainter than the large-scale northern survey. The repeated imaging, in addition to allowing the detection of fainter images, will also be crucial for the detection of variable objects.

A simulated redshift-space distribution of galaxies expected from the Sloan Survey in a 6° thick slice along the survey equator is presented in Fig. 1. This slice, based on a cold-dark-matter cosmological model simulation, contains approximately 6% of the galaxy redshift survey. The redshift histogram of galaxies in the simulated spectroscopic survey is shown in Fig. 2. The distribution peaks at z $\sim$ 0.1 - 0.15, with a long tail to z $\sim$ 0.4.

Several thousand rich clusters of galaxies will be identified in the survey, many with a large number of measured galaxy redshifts per cluster (from the complete spectroscopic survey). A summary of the expected cluster data is given in §3. The SDSS project is expected to take five years to complete. Currently under construction, it is expected that a test-year trial of the system will begin in 1996. More detailed information on the SDSS project are provided in York *et al*. (1993) and Gunn *et al*. (1993).

## 3. Clusters of Galaxies

Large scale surveys such as the SDSS will provide a much needed advance in the systematic study of clusters of galaxies, which is currently limited by the unavailability of modern, accurate, complete, and objectively selected catalogs of clusters, and by the limited photometric and redshift information for the catalogs that do exist. The three main catalogs of rich clusters available, those of Abell (1958), Zwicky *et al*. (1961-1968), and Abell *et al*. (1989) have been obtained from the Palomar Sky Survey and ESO Southern survey, using eye selection of clusters on photographic plates. Even with this inaccurate procedure, and with only a relatively small number of cluster redshifts (*cf*. Struble and Rood 1991), the catalogs have contributed significantly to our understanding of large scale structure. Recent advances include wide-angle catalogs selected by objective algorithms from digitized photographic plates (Dodd & MacGillivray 1986, Picard 1991, Dalton *et al*. 1992, Lumsden *et al*. 1992) and from a deep CCD survey covering six square degrees (Postman *et al*. 1994). The SDSS survey will cover a considerably larger area and will have a complete spectroscopic survey of the cluster galaxies.

Clusters are the largest virialized structures known, and they represent the high-density end of the large scale structure. Theoretical analyses (*e.g.* Shandarin & Zel'dovich 1989) predict that clusters should form at the intersections of large scale sheets and filaments, and it appears from existing data that this is at least qualitatively the case. Clusters provide useful tracers of large scale structure, but their intrinsic properties, the distribution of those properties within the cluster population, and the evolution of those properties provide important information for studies of galaxy and cluster evolution and the dependence of galaxy properties on their environment.

## 3.1 The Cluster Catalog

The digital photometric and redshift catalogs produced by the SDSS will permit the selection of clusters by well-defined, automated algorithms. From the $10^6$ galaxies in the spectroscopic sample, we can identify clusters by searching for density enhancements in redshift space. We can also identify clusters from the much larger photometric sample ($5\times10^7$ galaxies) by searching for density enhancements in position-magnitude-color space. We expect to find approximately 4000 clusters of galaxies with a redshift tail to z ~ 0.5; many of the clusters will have a large number of measured galaxy redshifts per cluster (hundreds of redshifts for nearby clusters), and at least one or two redshifts for the more distant clusters. Maps for two simulated clusters as expected to be seen in the 3-D survey at different redshifts are shown in Fig. 3. The large and complete sample of clusters can be used to investigate numerous topics in the study of clustering and large-scale structure. I briefly illustrate some of these topics below.

3.2 Clusters as Tracers of Large Scale Structure

Clusters are known to be highly efficient tracers of the large-scale structure. Studies of the cluster distribution have yielded important results even though the number of cluster redshifts in complete samples has been quite small, ~ 100 to 300 (for recent reviews see Bahcall 1988, 1993). The SDSS will put such studies on a much firmer basis because of the larger sample size (roughly an order of magnitude more redshifts), the objective identification methods, and the use of cluster-finding algorithms that are less subject to projection contamination.

Catalogs of nearby superclusters have been constructed using Abell clusters with known redshifts, revealing structures on scales as large as ~ $150h^{-1}$ Mpc (Bahcall and Soneira 1984). There are claims for still larger, ~ $300h^{-1}$ Mpc structures in the Abell cluster distribution (Tully 1987); these are still controversial (Postman *et al*. 1989). Existing galaxy redshift surveys probe a much smaller volume than do the cluster catalogs, but to the extent they can be compared, the superclusters identified from the cluster distribution match up well with structures in the galaxy distribution; the Perseus-Pisces supercluster (Haynes & Giovanelli 1988), and the Great Wall (Geller and Huchra 1989) are both clearly seen in the distribution of clusters and superclusters. The SDSS redshift survey will identify a complete sample of superclusters and will clarify the relation between the cluster distribution and the galaxy distribution. The larger cluster sample should yield definitive answers about structure at very large scales. Cluster peculiar velocities, obtained directly from distance indicators such as the $D_n - \sigma$ measurements and indirectly from the anisotropy of the cluster correlation function in redshift space, will provide information about the internal dynamics of superclusters and the mass distribution on large scales.

The high amplitude of the cluster correlation function, $\xi_{cc}(r)$, was one of the earliest pieces of statistical evidence for strong clustering on very large scales (Bahcall and Soniera 1983; Klypin and Kopylov 1983; Bahcall 1988; Postman *et al*. 1992). Kaiser (1984) suggested that the amplifications of $\xi_{cc}$ relative to the galaxy correlation function reflects the tendency of clusters to form near high peaks of the primordial density fluctuations; this idea later became the origin for theories of biased galaxy formation. With the SDSS we will have large numbers of clusters defined objectively from uniform and accurate photometric and spectroscopic data, allowing one to measure $\xi_{cc}$ on large scales free from systematic effects. We can extend the correlation analysis in several ways, examining in detail the richness dependence of $\xi_{cc}$ (Bahcall and West 1992) as well as the dependence of the cluster correlation on other cluster properties (e.g., velocity dispersion, morphology). The results can place strong constraints on cosmological models as shown in Fig. 4 (Bahcall and Cen 1992).

Another topic of considerable interest is the alignment of galaxies with their host clusters, and of clusters with their neighboring clusters, their host superclusters, and other large-scale structures such as sheets and filaments. While the alignment of cD galaxies with their clusters is relatively well established (Binggeli 1982), alignments at larger scales are still uncertain (Chincarini, Vettolani, and De Souze 1988). The SDSS will provide a very large sample for alignment studies, including accurate position angles and inclinations. The existence of large scale alignments may place interesting constraints on the origin of structure in various cosmological models (West *et al*. 1989).

### 3.3 Global Cluster Properties

The complete cluster survey will allow a detailed investigation of intrinsic cluster properties. From the 2-D and 3-D information, we will be able to determine accurately such properties as cluster richness, morphology, density and density profile, core radius, velocity dispersion profile, optical luminosity, and galaxy content (morphological fractions and cD galaxies), and to look for correlations between these properties. The cluster catalogs will be matched with the data in other bands, in particular the X-ray. This will allow a systematic detailed study of global cluster properties and their cross-correlation. Measurements of galaxy density and morphology, velocity dispersion and X-ray emission will shed light on the nature of the intracluster medium and its impact on the member galaxies, and on the relative distribution of baryonic and dark matter (*cf*. White *et al*. 1993). High surface mass density clusters will be candidates for gravitational lenses; we can search the photometric data for systematically distorted background galaxies, especially in the south, and target these clusters for yet deeper imaging surveys with larger telescopes. These lensing studies will allow mapping out the dark matter distribution within the clusters (Tyson *et al*. 1990, Kaiser & Squires 1993).

The large statistical sample of clusters will produce new insights into a number of issues associated with structure formation and evolution:

- We will be able to calculate velocity dispersions for ~ 1000 clusters with $z \leq 0.2$ compared to a few dozen clusters with reliable dispersion measurements currently available. For nearer clusters we will have well-sampled galaxy density profiles and velocity dispersion profiles; these allow careful cluster mass determinations. The distribution of cluster masses and velocity dispersion, i.e., the cluster mass-function and velocity-function will be determined accurately; these functions will place strong constraints on cosmological models as illustrated in Fig. 5 (Bahcall and Cen 1992; White *et al*. 1993).

- With optical luminosities, galaxy profiles, and velocity dispersion profiles, we can measure accurate mass-to-light ratios for a large sample of clusters. These provide constraints on $\Omega$ and the bias parameter of galaxies in clusters.

- The luminosity and velocity dispersion profiles provide information about the relative distributions of luminous and dark matter, and hence about the nature of the dark matter itself. More powerful constraints can be obtained when the optical data are combined with X-ray observations, or with dark matter maps derived from lensing of background galaxies. The ratio of baryonic mass to dark mass, combined with nucleosynthesis arguments, can yield a powerful constraint on $\Omega$ (White *et al*. 1993).

• The density and dispersion profiles also retain clues about the history of cluster formation. The frequency of subclustering and non-virial structures in clusters (Geller 1990) tells us whether clusters are dynamically old or young. In a gravitational instability model with $\Omega \approx 1$, structure continues to form today, while in a low-$\Omega$, open universe, clustering tends to freeze out at moderate redshift. Cluster profiles and substructure statistics thus provide a diagnostic for $\Omega$ which is independent from the direct measures of the mass density. The SDSS spectroscopic survey will provide an excellent data-base for such studies.

• The evolution of the cluster population provides a direct look at the history of galaxy clustering, thus providing another powerful diagnostic for $\Omega$. Current investigations are limited by the small numbers of known high-redshift clusters, and by the fact that high- and low-redshift clusters are selected in different ways. The uniform SDSS cluster sample will greatly improve the current situation. The number of clusters will be large; we will have redshifts for brightest cluster galaxies to $z \approx 0.5$, and for more distant clusters we can obtain fairly accurate estimated redshifts from the photometry alone (using the colors and apparent luminosity function of member galaxies). The southern photometric survey will probe the cluster population to redshifts above unity.

• In addition to the properties of clusters themselves, the properties of galaxies in clusters are also of great interest. It is well known that elliptical and S0 galaxies are more common in rich clusters than in the field, and that the fraction of early types depends on cluster density (Dressler 1980). It has recently been suggested, however, that a morphology-*radius* relation is a better description of existing data than a morphology-density relation (Whitmore *et al*. 1993). The SDSS will provide a powerful database for examining such issues, going beyond statistical characterization to detailed studies of the effects that might produce morphological segregation. The interaction of individual galaxies with the intracluster medium and tidal fields from their neighbors can be studies
in detail using the accurate morphological, photometric, and spectroscopic data for the
galaxies.

• There is much recent evidence for strong evolution of galaxies in clusters, from color
data (Butcher and Oemler 1984), spectroscopic data (Dressler and Gunn 1990), and
morphological data (Dressler *et al*. 1993). With its photometric data for cluster galaxies,
extending to $z > 1$ in the south, the SDSS will quantify the Butcher-Oemler effect in detail
for thousands of clusters, and the spectroscopic sample may itself be deep enough to
detect evolution in properties of cluster galaxies. The SDSS photometric cluster sample
will be ideal for follow-up spectroscopic observations on larger telescopes.

3.4 <u>Large-Scale Flows and the Hubble Constant from Clusters</u>

There is another important use of clusters in the classical field of cosmology. It has
been known for a long time that brightest cluster galaxies (BCGs) are good standard
candles (Sandage 1961). This was more recently confirmed in a detailed study of the
photometry of ~ 100 of these objects by Lauer and Postman (1994), showing that the
dispersion is as small as $0^m.35$ and can probably be reduced significantly by the use of
velocity dispersion data. The Hubble diagram given by the few thousands BCGs obtained
from the SDSS would test for nonuniformity in the Hubble flow out to redshifts
approaching 0.3, thus tracing the peculiar velocity field on large scales and constraining $\Omega$.
Distances to the nearby sample of clusters can be obtained by the Tully-Fisher (1977) and
$D_n - \sigma$ (Lynden-Bell *et al*. 1988) techniques, thus establishing the zero point of the BCG
Hubble diagram and inferring a global value of the Hubble constant.

## 4. **Large-Scale Structure**

4.1 <u>Power Spectrum and Correlation Function</u>

The most basic statistical measures of galaxy clustering are the power spectrum and
the two-point correlation function. With the SDSS, we can measure the angular correlation
function from the photometric survey, and the redshift-space power spectrum and correlation function directly from the redshift survey. The photometric survey will allow galaxy selection to be precise and uniform across the survey region, a critical requirement for accuracy.

The power spectrum of the galaxy distribution has been measured from existing redshift surveys out to scales of about $100 \, h^{-1}$ Mpc with approximately a factor of two precision (Vogeley *et al*. 1992; Fisher *et al*. 1993; Park *et al*. 1994). These scales are roughly one order of magnitude smaller than the smallest scales for which the COBE satellite has measured fluctuations in the cosmic microwave background (CMB) (Smoot *et al*. 1992). In the simplest interpretation, the CMB anisotropies are directly related to fluctuations in the gravitational potential (Sachs & Wolfe 1967). With the SDSS redshift survey, we will measure the galaxy power spectrum on scales where the Sachs-Wolfe effect is measurable in the CMB, and thus directly compare the amplitude of galaxy fluctuations to that of gravitational potential fluctuations.

The SDSS spectroscopic survey will allow measurement of the power spectrum over an enormous range of scales with unprecedented precision. This precision is illustrated in Fig. 6, which shows the power spectrum of the $\Omega h = 0.3$ cold dark matter model (which is roughly consistent with existing data), together with an estimate of the error bars we would find from the SDSS northern redshift sample. For comparison, the open points with larger error bars represent the power spectrum estimation of IRAS galaxies by Fisher *et al*. (1993). If the assumed power spectrum is a reasonable approximation of the truth, we will be able to measure it beyond the turnover scale, which corresponds to the radius of the horizon at matter-radiation equality, and out to scales probed by COBE. The detailed shape of the power spectrum reflects the physical source of the primordial fluctuations and the matter content of the universe. Current observations are sufficient to rule out some theoretical models, such as the simplest and most attractive versions of the cold dark matter scenario. The high-precision measurements afforded by the SDSS, especially near the peak of the power spectrum, will provide much more stringent tests of theoretical models.

The primordial power spectrum of mass fluctuations differs from the present-day power spectrum of the galaxy distribution because of (1) non-linear gravitational effects (which are strongest on small scales), (2) distortions by peculiar velocities in redshift space (which enhance the spectrum on large scales and depress it on small scales), and (3) possible "biasing" between the galaxy and mass distributions (with unknown scale dependence). These effects complicate the comparison between theory and observation, but they are important in their own right; they can be explored and can place limits on cosmological models, biasing and $\Omega$ (see below).

4.2 Redshift-Space Distortions

Peculiar velocities shift galaxies along the line of sight in redshift space, creating anisotropic distortions in the correlation function and the power spectrum. On small scales, velocity dispersions stretch dense structures along the line of sight. Thus clusters of galaxies appear as "fingers-of-God" in redshift space. On large scales, coherent outflows from voids and inflows to superclusters amplify density contrasts along the line of sight (Sargent & Turner 1977; Kaiser 1987). The distortions can be quantified by measuring the correlation function or power spectrum as a function of *direction* with respect to the line of sight, at fixed values of the pair distance or the wavelength. From the anisotropy of the correlation function, one can extract moments of the galaxy pairwise velocity distribution (Davis & Peebles 1983; Fisher *et al*. 1994). These provide powerful constraints on theoretical models, particularly on the values of $\Omega$ and the "bias parameter" $b$, which describes the relative amplitude of galaxy and mass fluctuations. On scales in the linear regime, distortions of the correlation function or the power spectrum depend directly on the combination $\Omega^{0.6}/b$ (Kaiser 1987; Hamilton 1992).

The SDSS will provide the large number of galaxies, large survey volume, and freedom from systematic errors that are needed to measure angular modulations of low-amplitude clustering on large angular scales. From the largest linear scales we will obtain precise measurements of $\Omega^{0.6}/b$ . By examining the transition from the linear regime to the non-linear regime, we can break the degeneracy between $\Omega$ and $b$ (Cole *et al*. 1994). Obtaining separate constraints on these parameters requires an enormous data set, and we expect that the SDSS will allow an unambiguous distinction between a high density universe with biased galaxy formation and a low-density universe in which galaxies trace mass. We will be able to test the robustness of this distinction by applying the same analysis to galaxies of different types − these may have different bias factors, but they should imply the same value of $\Omega$.

### 4.3  Topology of the Galaxy Distribution

A particularly interesting way to measure high-order clustering in the galaxy distribution is to examine the topology of high- and low-density regions. At the qualitative level, one would like to know whether the galaxy distribution is best described by a sponge topology, in which high- and low-density regions have equivalent connectivity, a bubble topology, in which voids are separated from each other by high-density walls, or an isolated cluster topology, with isolated high-density regions residing in a low-density sea. A quantitative measure of topology is the genus curve of Gott *et al*. (1986), which measures the genus of isodensity contour surfaces (the number of "holes" or "handles" minus the number of isolated regions) as a function of the fractional volume enclosed by the contour.

Because of its large volume, the SDSS redshift survey will provide a measurement of the topology on linear scales with sufficient accuracy to measure small deviations from the measurement of the small scale genus curve which has a factor of ten higher signal-to-noise ratio than previous measurements. These topological analyses will provide tight constraints on primordial fluctuations.

### 4.4  Clustering of Different Galaxy Types

One of the most exciting features of the SDSS is that we can study clustering in detail for different classes of galaxies. It is well known that the fraction of elliptical and S0 galaxies rises dramatically in the cores of rich clusters (Bahcall 1977; Dressler 1980; Postman and Geller 1984; Giovanelli *et al.* 1986). However, it is unclear whether this morphological segregation extends to large scales, and whether it occurs outside of the densest regions. One major obstacle in addressing these issues is the small size of existing samples. Random and systematic errors in the morphological types listed by existing catalogs also pose serious problems. With the SDSS we can study small and large scale differences in the clustering of galaxies of different morphological types, different colors, different luminosities, different surface brightnesses, and different spectroscopic properties. The SDSS photometric sample will be cross-correlated with samples from large-angle surveys in other bands, *e.g.* X-rays from the ROSAT all-sky survey, near infrared from the Two-Micron All Sky Survey, far-infrared from IRAS, and radio from the VLA. We can therefore compare the clustering of different type optical galaxies to that of galaxies that are bright in these various bands.

Clustering studies divided by galaxy type will be particularly valuable for addressing important clues about the physical processes that influence galaxy formation and determine galaxy properties. They will also provide crucial information about the relation between the galaxy and mass distributions, thereby helping to separate effects of biased galaxy formation from signatures of primordial fluctuations and cosmological parameters.

The clustering of quasars, investigated as a function of redshift to $z > 4$, will add a new dimension to our understanding of the large-scale structure: it will allow a detailed study of the *evolution* of large-scale structure from very early times to the present. This evolution will provide critical information for models of structure formation.

4.5  Distance Indicators and Large Scale Velocities

The high-resolution galaxy spectra for $10^6$ galaxies will provide linewidths and velocity dispersions of the galaxies. Combined with the photometric data, this information will allow one to derive distance indicators such as the Tully-Fisher and $D_n - \sigma$ methods and therefore study the peculiar velocity field out to very large scales. Two fundamental questions can be addressed using this data set: the rate at which flows, averaged over progressively larger scales, converge to the CMB rest-frame, and the relation between the mass-density field (inferred from the flow pattern within the survey region) and the galaxy-density field obtained from the redshift survey. The answers to these questions provide important information about the amplitude of mass fluctuations on very large scales and about the values of the density parameter and the bias parameter

We estimate that over 300 independent, statistically significant measures of the velocity field can be obtained across the full SDSS survey volume. Comparison between the galaxy number densities and mass density derived from the velocities yields a direct measure of the cosmological density parameter and the bias, in the combination $\Omega^{0.6}/b$ (*cf.* Dekel *et al.* 1993), a measure that is independent of that obtained from the anisotropy of clustering in redshift space discussed in section 4.2.

**Acknowledgement**

The SDSS is a collaborative effort of several institutions and a large number of scientists and engineers. The summary presented here is drawn from this collaboration and is greatfully acknowledged. The participating institutions are Fermilab, Institute for Advanced Study at Princeton, Japan Participation Group, Johns Hopkins University, Princeton University, University of Chicago, University of Washington, and the U.S. Naval Observatory. The project is managed by the Astrophysical Research Consortium. Major financial support for the SDSS is provided by the Alfred P. Sloan Foundation, the NSF, and the above institutions.

**Figure Captions**

Fig. 1  The redshift-space distribution of galaxies in a 6° thick slice along the survey equator from a large N-body cosmological simulation (cold-dark-matter with $\Omega h \sim 0.3$).  This slice contains approximately 6% of the $10^6$ galaxy redshift survey.

Fig. 2  The redshift histogram of galaxies in the simulated spectroscopic sample.  The histogram is computed in bins of $\Delta z = 0.01$.

Fig. 3  Two simulated rich clusters: a richness class R = 2 at z = 0.02 (e.g., Coma cluster) (left) and R = 5 at z = 0.2 (right).  All dots shown are redshift candidates for the spectroscopic survey.  In the distant cluster, the filled symbols are background objects.

Fig. 4  The correlation length of the cluster correlation function as a function of mean cluster separation, from observations and cosmological simulations (Bahcall and Cen 1992).

Fig. 5  The mass-function of clusters of galaxies.  Observations (points) (Bahcall and Cen 1993); cosmological simulations of different $\Omega h$ CDM models (lines) (Bahcall and Cen 1992).

Fig. 6  Estimate of the galaxy power spectrum.  The solid points and dashed line show the linear theory power spectrum of $\Omega h = 0.3$ CDM, with error bar estimates appropriate for the Sloan Survey.  Open points are from Fisher *et al*. (1993).  The scales probed by COBE are shown.

**References**

Abell, G.O. 1958, *ApJS*, **3**, 211.

Abell, G.O., Corwin, H.G., and Olowin, R.P. 1989, *ApJS*, **70**, 1.

Bahcall, N.A. 1977, *ARAA*, **15**, 505.

Bahcall, N.A. 1988, *ARAA*, **26**, 631.

Bahcall, N.A. 1993, *Proc. NAS*, USA, **90**, 4848.

Bahcall, N.A. and Cen, R. 1992, *ApJL*, **398**, L81.

Bahcall, N.A. and Soneira, R.M. 1983, *ApJ*, **270**, 20.

Bahcall, N.A. and Soneira, R.M. 1984, *ApJ*, **277**, 27.

Bahcall, N.A. and West, M.J. 1992, *ApJ*, **392**, 419.

Binggeli, B. 1982, *AA*, **107**, 338.

Butcher, H. and Oemler, A. 1984, *ApJ*, **285**, 426.

Chincarini, G., Vettolani, G., and de Souza, R. 1988, *AA*, **193**, 47.

Cole, S., Fisher, K.B. and Weinberg, D.H. 1994, *MNRAS*, **267**, *785.*

Dalton, G.B., Efstathiou, G. Maddox, S.J. and Sutherland, W.J. 1992, *ApJL*, **390**, L1.

Davis, M. and Peebles, P.J.E. 1983, *ApJ*, **267**, 465.

Dekel, A., Bertschinger, E. and Faber, S.M. 1990, *ApJ*, **364**, 349.

Dodd, R.J. and MacGillivray, H.T. 1986, *AJ*, **92**, 706.

Dressler, A. 1980, *ApJ*, **236**, 351.

Dressler, A. and Gunn, J.E. 1990, in *Evolution of the Universe of Galaxies, Astronomical Society of the Pacific Conference Series,* **10**, 200.

Dressler, A., Oemler, A., Gunn, J.E. and Butcher, H. 1993, *ApJL*, **404**, L45.

Fisher, K.B., Davis, M., Strauss, M.A., Yahil, A. and Huchra, J.P. 1993, *ApJ*, **402**, 42.

Fisher, K.B., Davis, M., Strauss, M.A., Yahil, A. and Huchra, J.P. 1994, *MNRAS*, **267**, 927.

Geller, M. 1990, in *Clusters of Galaxies: Space Telescope Symposium Series $4*, W. Oegerle, M. Fitchett, and L. Danly, eds. (Cambridge: Cambridge University Press).

Geller, M. and Huchra, J.P. 1989, *Science*, **246**, 897.

Giovanelli, R., Haynes, M.P. and Chincarini, G. 1986, *ApJ*, **300**, 77.

Gott, J.R., Dickinson, M. and Melott, A.L. 1986, *ApJ*, **306**, 341.

Gunn, J.E. and Knapp, G. 1993, in Proceedings of "Sky Surveys", APS Conference
  Series 43, ed. B.T. Sofia.

Hamilton, A.J.S. 1992, *ApJL*, **385**, L5.

Haynes, M.P. and Giovanelli, R. 1988, in *Large Scale Motions in the Universe: A Vatican
  Study Week*, edited by V.C. Rubin and G.V. Coyne, S.J. (Princeton: Princeton
  University Press) p.31.

Kaiser, N. 1984, *ApJL*, **284**, L49.

Kaiser, N. 1987, *MNRAS*, **227**, 1.

Kaiser, N. and Squires, G. 1993, *ApJ*, **404**, 441.

Klypin, A.A. and Kopylov, A.I. 1983, *Sov. Astron. Lett.*, **9**, 41.

Lauer, T.R. and Postman, M. 1994, *ApJ*, **425**, 418.

Lumsden, S.L., Nichol, R.C., Collins, C.A. and Guzzo, L. 1992, *MNRAS*, **258**, 1.

Lynden-Bell, D., Faber, S.M., Burstein, D., Davies, R.L., Dressler, A., Terlevich, R.J.
  and Weger, G.W. 1988, *ApJ*, **326**, 19.

Park, C., Vogeley, M.S., Geller, M.J. and Huchra, J.P. 1994, *ApJ*, in press.

Picard, A. 1991, PhD. Thesis, California Institute of Technology.

Postman, M.P. and Geller, M.J. 1984, *ApJ*, **281**, 95.

Postman, M.P., Gunn, J.E., Oke, J.B., Schneider, D.P. and Hoessel, J.G. 1994, in
  preparation.

Postman, M., Huchra, J.P. and Geller, M. 1992, *ApJ*, **384**, 404.

Postman, M., Spergel, D.N., Sutin, B. and Juszkiewicz, R. 1989, *ApJ*, **346**, 588.

Sachs, R.K. and Wolfe, A.M. 1967, *ApJ*, **147**, 73.

Sandage, A. 1961, *ApJ*, **133**, 355.

Sargent, W.W. and Turner, E.L. 1977, *ApJL*, **212**, L3.

Shandarin, S. and Zel'dovich, Ya 1989, *Rev. Mod. Phys.*, **61**, 185.

Smoot, G.R., Bennett, C.L., Kogut, A., Wright, E.L., Aymon, J., Boggess, N.W.,
Cheng, E.S., De Amici, G., Gukis, S. and Hauser, M.G. 1992, *ApJL*, **396**, L1.

Struble, M.F and Rood, H.J., *ApJS*, **77**, 363.

Tully, R.B. and Fisher, J.R. 1977, AA, **54**, 661.

Tully, R.B. 1987, *ApJ*, **323**, 1.

Tyson, J.A., Wenk, R.A. and VAldes, F. 1990, *ApJL*, **349**, L1.

Vogeley, M.S., Park, C., Geller, M.J. and Huchra, J.P. 1992, *ApJL*, **391**, L5.

West, M.J., Oemler, A. and Dekel, A. 1989, *ApJ*, **346**, 539.

White, S.D.M., Navarro, J.F., Evrard, A.E. and Frenk, C.S. 1993, *Nature*, **366**, 429.

Whitmore, B.C., Gilmore, D.M. and Jones, C. 1993, *ApJ*, **407**, 489.

York, D., Gunn, J.E., Kron, R., Bahcall, N.A., Bahcall, J.N., Feldman, P.D., Kent,
S.M., Knapp, G.R., Schneider, D., and Szalay, A. 1993, "A Digital Sky Survey of the
Northern Galactic Cap", NSF proposal

Zwicky, F. *et al*. 1961-1968, *Catalog of Galaxies and Clusters of Galaxies* (Pasadena:
California Institute of Technology).